# Two-dimensional short-range spin-spin correlations in the layered spin-3/2 maple leaf lattice antiferromagnet $Na_2Mn_3O_7$ with crystal stacking disorder

B. Saha, A. K. Bera[*] and S. M. Yusuf [*]

*Solid State Physics Division, Bhabha Atomic Research Centre, Mumbai 400085, India*
*and Homi Bhabha National Institute, Anushaktinagar, Mumbai 400094, India*

A. Hoser

*Helmholtz-Zentrum Berlin für Materialien und Energie, 14109 Berlin, Germany*

* corresponding authors: akbera@barc.gov.in, smyusuf@barc.gov.in

## ABSTRACT

We report the nature of magnetic structure, microscopic spin-spin correlations and their dependence on the underlying crystal structure of the geometrically frustrated layered spin-3/2 maple-leaf-lattice (MLL) antiferromagnet $Na_2Mn_3O_7$ by a comprehensive neutron diffraction study. Crystal structural studies by x-ray and neutron diffractions reveal that the MLL layers (constituted by $Mn_3O_7^{2-}$ units) are well separated by non-magnetic Na layers. The studies also conclude the presence of stacking faults (in-plane sliding of magnetic MLL layers) as well as a distortion in the MLL of $Mn^{4+}$. Temperature dependent magnetic susceptibility, heat capacity, and neutron diffraction data yield a short-range antiferromagnetic (AFM) ordering below ~ 100 K without a long-range magnetic ordering down to 1.5 K. The analysis of the diffuse magnetic neutron scattering patterns by reverse Monte Carlo method reveals 2D spin-spin correlations within the MLL layers. Additionally, we establish a relation between the correlation length of the short-range magnetic ordering with the stacking faults through a varying synthesis condition. The present study, therefore, explores a microscopic picture of the crystal- and spin-structures, as well as their correlation, hence, provides an experimental insight of the magnetic ordering in a MLL AFM. Further, we have outlined the formation of several 2D frustrated lattice geometry having triangular plaquettes, including the MLL, by crystal engineering of the triangular lattice and their role on the stabilization of multiple novel chiral spin states which opens up a door for study of novel chiral spin states.

# I. INTRODUCTION

Physics of low dimensional magnetic systems is an interesting and challenging problem, and has drawn vast attention in the field of both fundamental and applied magnetism [1-7]. Especially, geometrically frustrated quasi-two dimensional (2D) layered magnetic materials based on transition metal oxides are of current interest due to their novel magnetic properties including various chiral ordered states [1,6] and spin disordered states. Frustrated spin systems often have highly degenerate magnetic ground states, and consequently a suppression of long-range magnetic ordering [7-10]. The magnetic properties of such frustrated 2D spin systems depend highly on the underlying lattice geometry. The competing nearest-neighbour exchange interactions provide frustration for triangular and kagome lattices; whereas, square and honeycomb lattices reveal frustration only under the presence of competing nearest-neighbour and next-nearest-neighbour exchange interactions. Further, the magnetic properties of frustrated spin systems can be tuned/modified by several means, such as introducing regular depletion/removal of magnetic atoms, substitution of nonmagnetic atoms, and/or by introducing distortion in the lattice.

In this regard, 2D maple-leaf-lattice (MLL)[11-13], a new class of geometrically frustrated spin system, is of present interest. The MLL is a naturally formed 2D lattice, where every $7^{th}$ lattice point is missing from a regular triangular lattice. The coordination number of the MLL ($z=5$) is intermediate to the triangular ($z=6$) and kagome ($z=4$) lattices. The kagome lattice is formed by periodic removal of the $1/4^{th}$ of lattice points from a regular triangular lattice. In case of a triangular lattice antiferromagnet with Heisenberg spins, the classical magnetic ground state involves a unique staggered vector chiral (SVC) spin structure, where vector chirality is opposite on two neighbouring triangles. In the case of Kagome lattice, there are two possible classical ground state spin configurations; (i) positive vector chiral (PVC) spin-structure known as "$q=0$" structure, and (ii) SVC, known as "$q=\sqrt{3}\times\sqrt{3}$" structure [14]. An ideal MLL with all uniform exchange interactions reveals a staggered vector chiral [SVC(1)] spin structure [11,15] where three spins of every third triangle (triangular plaquette) in the hexagonal ring of MLL form a 120° structure. Most interestingly, for a distorted MLL (keeping the six-fold symmetry invariant) with three non-equivalent AFM exchange interactions $J_d$, $J_t$, and $J_h$ (where $J_d \gg J_t > J_h$) three degenerated chiral spin structures, viz., (i) PVC, (ii) negative vector chiral (NVC), and (iii) a new staggered vector chiral [SVC(2)] spin-structures were reported depending on the relative strengths of $J$s [12]. In this kind of distorted MLL (with three non-equivalent exchange interactions), each of the six triangular plaquettes are made of three equivalent exchange interactions $J_t$, and such triangular plaquettes are connected by unequal exchange interactions $J_d$ (one number) and $J_h$ (two numbers). However, MLL magnets are rarely explored experimentally due to unavailability of suitable real compounds. In this regard, the new layered compound $Na_2Mn_3O_7$ (with alternative stacking of magnetic $Mn_3O_7^{2-}$ and nonmagnetic $Na^+$ layers) involving distorted MLL of $Mn^{4+}$ ions (spin-3/2) is of current interest. The compound $Na_2Mn_3O_7$ has been recently investigated in great detail for Na-ion battery application, and reported to be a highly efficient battery material [16-24]. However, there is only one report in the literature on the magnetic properties of $Na_2Mn_3O_7$ based on NMR experimental results [25], which claimed a nonmagnetic ground state. Considering the 2D MLL of spin 3/2 in $Na_2Mn_3O_7$ and theoretical predictions of long-range classical magnetic ground states [11,12], a non-magnetic ground state is very unusual and debatable. It, therefore, demands a direct determination of microscopic magnetic spin-spin correlations by a probe like neutron diffraction to elucidate the true nature of magnetic ground state of this compound involving a fascinating 2D magnetic MLL lattice.

In the present paper, by performing a comprehensive neutron diffraction study and data analysis using the reverse Monte Carlo (RMC) method, we have determined the microscopic magnetic spin-spin correlations, and their temperature evolution in the spin-3/2 MLL compound $Na_2Mn_3O_7$. We have further demonstrated the dependence of the spin-spin correlation length of the magnetic ordering on the crystal structural stacking faults intrinsic to $Na_2Mn_3O_7$. The nature of magnetic ground state spin structure for $Na_2Mn_3O_7$, having distorted MLL with nine non-equivalent nearest neighbour exchange interactions, has been outlined. Besides, we have provided a detailed discussion on the realization of several 2D geometrical frustrated lattices based on triangular plaquettes, including the MLL, by crystal engineering (depletion of lattice points and bonds) of the triangular lattice, and the roles of such underlying crystal lattice geometries on the stabilization of various novel chiral spin structures.

## II. EXPERIMENTAL DETAILS

The polycrystalline samples of $Na_2Mn_3O_7$ were prepared by solid-state synthesis method. Stoichiometric mixtures of high purity precursors (>99.99%) $NaNO_3$ and $MnCO_3$ were thoroughly ground by using an agate mortar and pestle. The mixture was then heated at 600 °C for 4 hours in an alumina crucible under an oxygen flow environment. To investigate the role of stacking faults on spin-spin correlations, additional samples were synthesized with annealing times of 14 and 28 hours.

Room temperature powder x-ray diffraction study (using a laboratory source x-ray diffractometer) was carried out by using a Cu-$K_\alpha$ radiation over the scattering wave vector range of $0.7 Å^{-1} \leq Q \leq 5.75 Å^{-1}$.

Temperature dependent dc magnetization measurements were performed on the 4 hours annealed sample by employing a commercial vibrating sample magnetometer (VSM) (Cryogenic Co. Ltd., UK).

To carry out the experiments, the sample was put inside a straw and mounted on the VSM. The temperature dependent dc magnetization measurements were carried out in the warming cycle over the temperature range of 1.5-300 K, after cooling the sample in the absence of field (ZFC condition). Isothermal magnetization measurements were carried at 2 K over the field range ±90 kOe after cooling the sample in the ZFC condition.

Neutron diffraction measurements on all three samples (annealed for 4, 14, and 28 hours) were carried out at 5, 100 and 300 K using the powder diffractometer PD-I ($\lambda$=1.094 Å) at Dhruva research reactor, Bhabha Atomic Research Centre, Mumbai, India. The polycrystalline samples were filled within an 8 mm cylindrical vanadium container and attached to the tail of the closed cycle helium refrigerator of the instrument for the measurements. Additional, detailed temperature dependent neutron diffraction measurements over 1.8-165 K were carried out on the 4 hours annealed sample by using the powder diffractometer E6 ($\lambda$=2.4 Å), Helmholtz-Zentrum Berlin, Berlin, Germany. The magnetic diffuse neutron scattering patterns were analysed by the RMC method using the *"SPINVERT"* software [26].

## III.  RESULTS AND DISCUSSIONS:

*A.    Crystal Structural Correlations:*

The room temperature x-ray diffraction pattern of the polycrystalline samples $Na_2Mn_3O_7$ (annealed for 4 hours) [Fig. 1(a)] shows the presence of Bragg and diffuse peaks as reported earlier by several groups [19,21,27], indicating the presence of stacking faults in the layered crystal structure of $Na_2Mn_3O_7$ [Fig. 1(b-c)]. The crystal structure is composed of alternating stacking of $Mn_3O_7^{2-}$ layers and Na-ion layers along the [1-10] direction [16]. The derived lattice parameters $a$ = 6.656(5) Å, $b$ = 6.835(1) Å, $c$ = 7.522(2) Å, $\alpha$ = 106.27(1)°, $\beta$ = 106.62(2)°, and $\gamma$ = 111.71(2)° are in good agreement with those reported earlier. The measured neutron diffraction pattern (green scattered points) [Fig. 1(d)] also reveals broadening of selective Bragg peaks in comparison with the same (red line) expected for a stacking fault free crystal structure. The analysis of the neutron diffraction pattern by using the *"Faults"* software [28] reveals substantial stacking faults where the in-plane sliding of ~ 25 % of the magnetic layers are found. The faulted layers are shifted by ~0.35$a$, 0.34$b$, and 0.31$c$ from their desired position in the triclinic crystal structure. The simulated pattern considering the stacking faults is shown by the blue line in Fig. 1(d). Further, the MLL of the $Mn_3O_7^{2-}$ layers is distorted and consists of nine non-equivalent Mn-Mn direct distances [a detailed discussion is given later in the neutron diffraction study section] which leads to nine non-equivalent magnetic exchange interactions.

## B.     Bulk Magnetic Properties:

The temperature-dependent dc magnetic susceptibility $\chi(T)$ curve under an applied field of 10 kOe is shown in Fig. 2(a). The curve shows a broad peak centred at ~115 K indicating the onset of a 2D short-range antiferromagnetic (AFM) order as expected from the layered structure of $Na_2Mn_3O_7$. Upon further lowering of the temperature the magnetic susceptibility $\chi(T)$ curve shows a Curie-like upturn below ~10 K due to the presence of some isolated $Mn^{4+}$ spins [29-34] because of crystal structural disorders. The nature of the susceptibility curve is similar to that reported earlier [25]. Further, as reported earlier the temperature dependent heat capacity curve does not show any anomaly down to 2 K suggesting [Fig. 2(a)] an absence of long-range magnetic ordering.  Besides, the $\chi T$ vs $T$ curve [inset of Fig. 2(a)] shows a deviation from a constant value (expected for a paramagnetic state) below ~560 K. With decreasing temperature, the $\chi T$ vs $T$ curve decreases sharply below ~150 K where 2D short-range magnetic correlations have been confirmed by neutron diffraction study (discussed later).

Further, we have analysed the magnetic susceptibility data by a high temperature series expansion (HTSE) method. For a 2D spin system, the magnetic susceptibility at high temperatures can be represented by a power series as [35]:

$$\chi = \chi_0 + (Ng^2\mu_B^2 S(S+1)/3k_B T) \sum_{l=1}^{7} 1 + a_l (J/kT)^l \qquad (1)$$

where, $N$ is Avogadro number, $k_B$ is the Boltzman constant, $a_l$'s are coefficients, $\chi_0$ represents the contributions from the temperature independent Van Vleck paramagnetic term and diamagnetic term due to atomic cores. As the values of coefficients $a_l$ for a MLL lattice are unknown, we have considered the values $a_1$=4, $a_2$=3.2, $a_3$=-2.186, $a_4$=0.08, $a_5$=3.45, $a_6$=-3.99 that were reported for the closest triangular lattice [35].  Eq. (1) reproduces the experimentally measured $\chi(T)$ curve over 80-500 K with a single variable of exchange constant $J/k_B$ = -17.52(2) K [red line in Fig. 2(b)], thus, suggesting a 2D nature of the magnetic lattice. Whereas, the earlier report on $Na_2Mn_3O_7$ claimed that the magnetic susceptibility curve can be explained by a 1D Heisenberg spin-chain model [25]. For a comparison, we have also fitted the susceptibility curve [blue dotted line in Fig. 2(b)] by the 1D Heisenberg spin chains model [36]:

$$\chi = \chi_0 + (Ng^2\mu_B^2 S(S+1)/3k_B T) \frac{\sum_{l=1}^{7} 1+a_l(J/kT)^l}{\sum_{l=1}^{7} 1+b_l(J/kT)^l} \qquad (2)$$

where $a_1$=0.48166, $a_2$=0.91348, $a_3$=0.11791, $a_4$=0.0054266, $a_5$=0.00000065501, $b_1$=2.9834, $b_2$=5.8136, $b_3$=7.3541, $b_4$=2.2637, and $b_5$=0.12999 [36].  However, it is evident that the 2D magnetic lattice model provides a better agreement (agreement factor, $R^2$ =0.997) with the experimental data as compared to that for the 1D model ($R^2$ =0.978). Here we would like to point out that our neutron diffraction study (discussed below) reveals 2D spin-spin correlations at low temperatures.

The isothermal magnetization $M(H)$ curve, measured at 2 K, shows almost a linear behaviour without any saturation even up to 90 kOe [Fig. 3]. The value of magnetization $M$ ~0.06 $\mu_B$/f.u. under 90 kOe is negligibly small (~2 %) as compared to the expected saturation magnetization value of 3 $\mu_B$/f.u. (=$gS$=2×3/2 for $Mn^{4+}$ ions) indicating a predominant antiferromagnetic spin structure. However, a weak hysteresis (remanent magnetization $M_R$ ~0.0041 $\mu_B$/f.u., and coercive field, $H_C$ = 470 Oe) is observed, possibly due to a spin canting. A field induced transition at ~ 30 kOe is also evident which is better seen in the $dM/dH$ curve [bottom right inset of Fig. 3].

### C. *Neutron Diffraction and Short-range Magnetic Correlations:*

To understand the nature of microscopic spin-spin correlations low temperature high resolution neutron diffraction measurements were carried out down to 1.8 K. Low temperature neutron diffraction patterns, measured on E6 instrument (below 90 K) [Fig. 4], reveal a broad diffuse magnetic peak over $Q$ ~ 1.2-2.5 Å$^{-1}$ corresponding to short-range spin-spin correlations. The absence of magnetic Bragg peak down to 1.8 K confirms the absence of a long-range magnetic ordering in $Na_2Mn_3O_7$. Similar broad diffuse magnetic peaks in the neutron diffraction patterns were reported for several quasi-2D magnetic systems [6,37-41]. The profile of the broad peak depends on the dimensionality of the magnetic ordering where a symmetric/asymmetric peak is expected from a 3D/2D ordering, respectively. In the present case, the observed asymmetric sawtooth type peak shape of the diffuse magnetic scattering [Fig. 4(b)] is attributed to 2D short-range magnetic correlations [37-41].

To determine the nature of spin-spin correlations of the short-range magnetic ordering, the diffuse magnetic diffraction patterns were analysed by the RMC method using the *"SPINVERT"* computational program [26] which was successfully employed for several low-dimensional spin systems [42-44]. As the *"SPINVERT"* program works on an orthogonal unit cell, we have transformed the triclinic unit cell of $Na_2Mn_3O_7$ to an orthorhombic unit cell [Fig. 5] by a projection method. The transformed orthorhombic unit cell lattice parameters are $a=\sqrt{(21)}p$ =13.0775 Å, $b=\sqrt{(7)}p$=7.5242 Å, $c=2d$=11.14 Å, and $\alpha=\beta=\gamma=90°$, where $p$ is the distance between the nearest neighbor $Mn^{4+}$ ions within the planes and $d$ is the separation between the $Mn_3O_7^{2-}$ layers, respectively. We have performed the RMC analysis by using a super cell 10×10×6 (13800 spins) of the orthorhombic unit cell. Initially, a random spin configuration was considered where all the spins were fixed to lattice points of the magnetic sites. During the refinement, only the orientations of the spins were varied to fit the experimental magnetic scattering pattern [45-47]. For each temperature, 15 independent refinements were performed. An average of the outputs was considered for the spin-spin correlations and scattering patterns in the reciprocal planes. For each of the refinements total 500 moves per spin were considered.

The experimentally measured magnetic diffuse scattering patterns at 1.8, 17, and 65 K along with the fitted curves are shown in Fig. 6(a-c). The corresponding spin-configuration solutions were used to regenerate the magnetic scatterings profiles on selected scattering planes ($hk0$), ($h0l$) and ($0kl$) [Fig. 6(d-l)] by using the "*SPINDIFF*" program [26]. Within the ($hk0$) plane (magnetic MLL plane), the symmetric types of scatterings reveal an isotropic magnetic spin-spin correlation. On the other hand, within the ($h0l$) and ($0kl$) planes, rodlike diffuse scatterings along the $l$ axis are evident [Fig. 6(e-f)]. Such rodlike scatterings are characteristic of 2D magnetic correlations within the ($hk0$) plane i.e., within the magnetic layers of $Na_2Mn_3O_7$. The rodlike scatterings become sharper with decreasing temperature revealing an increase of the spin-spin correlation length [Fig. 6(d-l)]. The 2D magnetic correlations are consistent with the layered crystal structure of $Na_2Mn_3O_7$ that provides stronger in-plane interactions (through Mn-O-Mn superexchange interaction pathways). On the other hand, the large separation (5.57 Å) between magnetic layers, by nonmagnetic Na layer, provides much weaker interactions between the MLL planes.

Further, we have calculated real space spin-spin correlation function $\langle \vec{S}(0).\vec{S}(r) \rangle$ from the refined spin configurations by using the "*SPINCORREL*" program [26]. A pictorial representation of the spin-configuration solution obtained from the RMC fitting of the powder magnetic diffuse scattering pattern at 1.8 K is shown in Fig. 7(a) [26]. The real space spin-spin correlations $\langle \vec{S}(0).\vec{S}(r) \rangle$ are calculated for each of the spin-configurations obtained from the 15 independent RMC refinements, and then taken an average over all of them for the presentation in Fig. 7(b). The average $\langle \vec{S}(0).\vec{S}(r) \rangle$ (where, $r$ signifies the distance between two magnetic sites in $Na_2Mn_3O_7$) [Fig. 7(b)] at 1.8 K reveals a ferromagnetic nearest-neighbour (NN) ($r_{avg}$ ~ 2.8796 Å), and antiferromagnetic next-nearest-neighbour (NNN) ($r_{avg}$ ~ 4.9053 Å) and next-to-next nearest-neighbour (NNNN) ($r_{avg}$ ~ 5.6908 Å) in-plane spin-spin correlations. It is further evident from Fig. 7(b) that the $\langle \vec{S}(0).\vec{S}(r) \rangle$ almost vanishes at ~8 Å. An increase of the in-plane NN, NNN and NNNN spin-pair correlations [Fig. 7(c)] is found with decreasing temperature.

Now we shed light on the magnetic spin structure of $Na_2Mn_3O_7$. The ideal MLL with all uniform NN AFM exchange interactions leads to 2D long-range SVC(1) spin structure [11,12]. With the introduction of lattice distortion in MLL (keeping the six-fold symmetry invariant), three non-equivalent AFM exchange interactions $J_d$, $J_t$ and $J_h$ appear [12]. In this case, three long-range chiral spin structures (i) PVC, (ii) NVC and (iii) SVC(2) can be stabilized [12] depending on the relative strengths of the exchange interactions. Whereas, for the studied compound $Na_2Mn_3O_7$, the MLL is further distorted and contains nine non-equivalent Mn-Mn direct distances (2.726-2.974 Å) [Fig. 8(a) and Table I] and consequently, nine non-equivalent exchange interactions $J_{d1}$, $J_{d2}$, $J_{d3}$, $J_{t1}$, $J_{t2}$, $J_{t3}$, $J_{h1}$, $J_{h2}$, and $J_{h3}$ (notations are considered to compare with that reported in [12]). This leads to a two-fold symmetry for the studied compound $Na_2Mn_3O_7$. For the distorted MLL in $Na_2Mn_3O_7$, the possible magnetic ground state spin structures can be qualitatively derived by considering the geometrical arrangement of the exchange interactions. In order to qualitatively consider the sign of the exchange interactions, we follow the report by Haraguchi *et al.*, [12] where it is reported that the sign of the $Mn^{4+}$-$Mn^{4+}$ coupling (for a pair of edge sharing $MnO_6$ octahedra) depends strongly on the direct Mn-Mn distances. An AFM direct-exchange interaction is predicted to be dominated for a distance, $d_{Mn-Mn}$ < 2.85 Å. Whereas, a FM superexchange interaction is predicted to be dominated for a distance, $d_{Mn-Mn}$ > 2.85 Å [12]. By considering the above concepts the exchange couplings $J_{h1}$ and $J_{h2}$ in $Na_2Mn_3O_7$ involving the Mn-Mn distances of 2.783 and 2.726 Å, respectively, are expected to be AFM. All the other seven interactions, involving Mn-Mn direct distances longer than 2.85 Å, are expected to be a FM. Considering two AFM ($J_{h1}$ and $J_{h2}$) and seven FM ($J_{d1}$, $J_{d2}$, $J_{d3}$, $J_{t1}$, $J_{t2}$, $J_{t3}$, and $J_{h3}$) exchange interactions, there are four possible collinear 2D long-range AFM structures [DDDD/UUUU, DDUU/UUDD, DUDU/UDUD, and DUUD/UDDU] [Fig. 8(b-e)]. The notation D/U represents the central hexagonal ring of the MLL having four (out-of-six) down/up spins, respectively. All the magnetic ground state spin structures are derived by considering the non-frustrated spin arrangement on the central hexagonal ring which consists of the AFM exchange interactions $J_{h1}$ and $J_{h2}$, and the FM exchange interactions $J_{h3}$. Here, we would like to mention that a 120° chiral magnetic structure can be possible in the distorted MLL with the two-fold symmetry, like in the case of $Na_2Mn_3O_7$, only when all the exchange interactions ($J_t$'s) within the $J_t$ triangular plaquettes (shaded triangles) are AFM and equal in strength.

To compare with the experimentally measured pattern, we have simulated the powder magnetic neutron diffraction patterns for the four collinear 2D long-range AFM structures [Fig. 8(f)] with ferromagnetic inter planar correlation. In order to compare with the experimentally observed diffuse magnetic pattern of $Na_2Mn_3O_7$ [Fig. 8(h)], the simulated magnetic neutron diffraction patterns [Fig. 8(f)] are regenerated considering short-range spin-spin correlations in the MLL plane [Fig. 8(g)]. It is evident that the simulated magnetic neutron diffraction patterns are different from the observed pattern for $Na_2Mn_3O_7$. This clearly indicates that the real spin configuration in $Na_2Mn_3O_7$ is non collinear and more complex as evident from Fig. 7(a). The observed non-collinear magnetic structure differs from the chiral magnetic structure with 120° spin arrangements as predicted theoretically for a MLL involving equivalent exchange interactions within triangular plaquettes. The present distorted MLL, which possesses 2-fold lattice symmetry, does not allow the 120° spin arrangements. On the other end, a collinear magnetic structure is expected for an extremely distorted MLL due to lifting of magnetic frustration. The present compound $Na_2Mn_3O_7$, thus, refers to a frustrated spin system intermediate to the above two cases. Further, we would like to point out that the observed 2D short-range magnetic correlations in $Na_2Mn_3O_7$ containing strong stacking faults (i.e., displacement of magnetic layers which is described in crystal structural section) is in sharp contrast to a 2D long-range magnetic ordering predicted theoretically for a MLL [11,15]. Therefore, the stacking faults not only affect the spin-spin correlations between the magnetic planes, but also have a strong role on the in-plane spin-spin correlations.

In order to understand the effect of varying stacking faults on the magnetic correlations, we also synthesized $Na_2Mn_3O_7$ samples with annealing times of 14 and 28 hours. Powder neutron diffraction patterns (measured at 300 K using the PD-I) of the $Na_2Mn_3O_7$ samples prepared with different annealing times (4, 14 and 28 hours) are shown in the left column of Fig 9(a). With increasing the annealing time, the neutron diffraction patterns reveal an increase in the intensity of the Bragg peaks (012), (102) and (-203), which are affected by stacking faults. The above observations suggest a reduction of stacking faults with the increasing annealing time [27]. Besides, with increasing annealing time the intensity of the diffuse magnetic peak (at $Q$ value 1.5 Å$^{-1}$) increases and the peak width becomes sharper [middle column of Fig 9(b)] revealing an increase in the spin-spin correlation length. We have analysed the patterns by the RMC method. The variations of $\langle \vec{S}(0).\vec{S}(r) \rangle$ [Fig. 9(c)], obtained from the RMC analyses, reveal an increase in the spin-spin correlations as well as the correlation length [Fig 9(d)]. This, therefore, indicates that a 2D long-range magnetic order, as predicted theoretically, may appear for a fault free sample of $Na_2Mn_3O_7$.

We now discuss the decisive role of the underlying crystal lattice geometry on the chiral magnetic ground state spin structures of 2D frustrated lattices. Chiral spin structures are one of the characteristic features of triangular lattice based geometrically frustrated spin systems. For an ideal triangular lattice (where triangular plaquettes share edges with its three neighbours) antiferromagnet with three-fold symmetry, and XY and Heisenberg spins, theoretical studies show that a long-range 120° spin structure is stabilized in the classical ground state [1]. Such 120° SVC spin structure is realized experimentally for several compounds, such as $KFe(PO_3F)_2$ [48], $AFe(SO_4)_2$ ($A$= Rb and Cs) [49], and $AFe(MoO_4)_2$ ($A$ = K, Rb, and Cs) [50]. In this spin structure, the vector chirality is opposite to the neighbouring triangular plaquettes and is known as a staggered vector chiral or SVC state [Fig. 10(a)]. In the case of a kagome lattice [Fig. 10(b-c)], formed by periodic removal of the 1/4$^{th}$ lattice points from a regular triangular lattice, the triangular plaquettes are connected by sharing their corners and having a six-fold symmetry. Such a change of lattice geometry allows the possibility of two classical vector chiral structures (i) SVC (or known as "$\sqrt{3}\times\sqrt{3}$" state) and (ii) PVC (or known as "$q=0$" state) [14], in contrast to the triangular lattice which allows only SVC structure. The PVC and SVC magnetic ground state spin structures are predicted theoretically [51] as well as realized experimentally for several kagome lattice compounds, such as, $KFe_3(OH)_6(SO_4)_2$ for PVC [14,52] and $ZnCu_3(OH)_6Cl_2$ [53] for SVC spin structures, respectively [14,53]. On the other hand, for a MLL, formed by a periodic removal of the $1/7^{th}$ ions from a regular triangular lattice, the triangular plaquettes are connected by sharing their edges with two neighbouring triangles revealing a six-fold symmetry [Fig. 10(d)], in comparison to the three triangles in the case of a triangular lattice. The classical ground state of an ideal MLL (i.e., without any structural distortion) was predicted to be a staggered vector chiral [SVC(1)] state where three spins of every third triangle in the hexagonal ring of MLL form a 120° structure [Fig. 10(d)], as in the case of triangular lattice [11,12,15]. The vector chirality on such every third triangle changes its sign alternately leading to the SVC(1) state for an ideal MLL.

Lattice distortion plays a significant role on the magnetic ground state properties. For a distorted triangular lattice with two in-plane NN exchange interactions $J_1$ and $J_2$, two different types of ordered magnetic phases, i.e., the SVC spin state and a collinear AFM spin state for $J_2/J_1 < 0.5$ and $J_2/J_1 > 0.5$, respectively, occur [54]. The presence of both chiral and collinear magnetic ground states was experimentally reported for a distorted triangular lattice compound $KFe(MoO_4)_2$ [55]. For a distorted kagome lattice (i.e., breathing kagome lattice), having two in-plane NN exchange interactions $J_1$ and $J_2$ (having three-fold symmetry) the magnetic ground state remains as SVC spin state over a wide range of $J_2/J_1$ and then becomes a spin-nematic state with weakly coupled local dimers and trimers [56,57]. Further, for the distorted kagome lattice with a broken three-fold symmetry, involving two non-equivalent AFM NN exchange interactions ($J_1$ and $J_2$), theoretical calculations predicted a transition from SVC spin structure to a coplanar stripe chiral (SC) spin structure for $J_2/J_1 >1$ [58]. Further, a theoretical study on a distorted kagome lattice with two non-equivalent NN exchange interactions $J_1$ and $J_2$ with opposite signs (one of them is FM, and the other one is AFM) reported a possibility of two degenerate collinear ground state spin structures [59]. With the introduction of further lattice distortion in kagome lattice involving three non-equivalent NN exchange interactions $J_1$, $J_2$ and $J_3$, a phase diagram involving a collinear $Q = 0$ magnetic state, two non-collinear coplanar $Q = (1/3,1/3)$ magnetic states, and a spin liquid state is predicted depending on the relative strengths of $J_1$, $J_2$ and $J_3$ [60]. On the other hand, for MLL, a lattice distortion involving three non-equivalent exchange interactions (keeping the three-fold symmetry invariant) can lead to three new vector chiral spin states (i) PVC, (ii) NVC, and (iii) SVC(2) states [12]. For the PVC and NVC states, the sign of the chirality (either right-handed, or left-handed) on every third triangle for each of these two states is uniform. In the case of SVC(2) state, the vector chirality on every third triangle changes its sign alternately like the SVC(1) state but the relative spin direction between such triangles differs by 90°. Further distortion in the MLL leads to two-fold symmetry and can significantly change the nature of spin-structure as evident in the present study. This anticipates future theoretical and experimental studies to explore the spin-spin correlation in the highly distorted MLL. The above discussion depicts how the periodic removal of lattice points from a regular triangular lattice results into different 2D geometrically frustrated lattices i.e., kagome lattice and MLL with different arrangements of the triangular plaquettes. Such change in the lattice geometry consequently leads to the possibility of multiple chiral spin structures, viz., chiral spin structures PVC, SVC and SC for kagome; and PVC, NVC and SVC for MLL. Thus as an outlook, the investigations of other possible 2D frustrated lattices based on triangular plaquettes that can be obtained from the triangular lattice by crystal engineering, such as, $1/4^{th}$ site depleted triangular lattice, $1/6^{th}$ site depleted triangular lattice, cube tile lattice, $1/6^{th}$ bond depleted triangular lattice [61] would be of immense interest for exploring the novel chiral spin structures as well as their correlations with the underlying lattice geometry.

## IV. SUMMARY AND CONCLUSIONS

Magnetic ordering, spin-spin correlations and their relation with the underlying crystal structure of the geometrically frustrated layered MLL spin-3/2 antiferromagnet $Na_2Mn_3O_7$ have been divulged through temperature dependent neutron diffraction, dc magnetization, and heat capacity studies. The crystal structure of $Na_2Mn_3O_7$ consists of an alternating stacking of $Mn_3O_7^{2-}$ and nonmagnetic Na-ion layers along the crystallographic [1-10] direction. Our analysis of neutron diffraction patterns reveals the presence of crystal structural stacking faults (in-plane sliding) of $Mn_3O_7^{2-}$ layers. The analysis of the neutron diffraction patterns further reveals a distortion in the MLL of $Mn^{4+}$ ions which leads to nine non-equivalent nearest neighbour distances. Bulk dc magnetization and heat capacity data reveal an absence of long-range magnetic ordering down to 2 K (lowest measured temperature). Low temperature neutron diffraction patterns show the appearance of magnetic diffuse scattering due to short-range magnetic ordering. Our comprehensive RMC analysis of the magnetic diffuse scattering patterns reveals 2D spin-spin correlations within the magnetic layers. We point out here that the observed 2D spin-spin correlations rule out the possibility of nonmagnetic singlet ground state in $Na_2Mn_3O_7$ as reported earlier by NMR study. The observed 2D spin-spin correlations are signatures for an intrinsic layered spin system, like $Na_2Mn_3O_7$. We have further shown that the spin-spin correlation length ($\xi$) can be enhanced by a reduction of stacking faults in the crystal structure through annealing over longer durations. The present study thus provides an experimental insight of the spin-spin correlations of the frustrated MLL spin-system and highlights its rich physics in view of geometrically frustrated spin systems. Besides, our in-depth discussion highlights the role of underlying lattice geometry of 2D frustrated lattices having triangular plaquettes on the occurrence of novel chiral spin structures and opens up a door for study of novel chiral spin structures.

## ACKNOWLEDGMENTS

The authors would like to acknowledge the help provided by V. B. Jayakrishnan, and Dr. M. Mukadam for the x-ray diffraction and specific heat measurements, respectively. B. Saha thanks the Department of Science and Technology (DST), Government of India, for providing the INSPIRE fellowship (Reference No. DST/INSPIRE/03/2017/000817, INSPIRE Grant No. IF180105).

## References

[1] H. Diep, *Frustrated spin systems* (World Scientific, 2013).
[2] S. Hazra, A. K. Bera, S. Chatterjee, A. Roy, and S. M. Yusuf, Spin-liquid signatures in geometrically frustrated layered kagome compounds $YBaCo_{4-x}Fe_xO_{7+\delta}$, Phys. Rev. B **104**, 144418 (2021).
[3] P. Suresh, K. Vijaya Laxmi, A. K. Bera, S. M. Yusuf, B. L. Chittari, J. Jung, and A. Kumar, Magnetic ground state of the multiferroic hexagonal $LuFeO_3$, Phys. Rev. B **97**, 184419 (2018).
[4] A. Yogi, A. K. Bera, A. Maurya, R. Kulkarni, S. M. Yusuf, A. Hoser, A. A. Tsirlin, and A. Thamizhavel, Stripe order on the spin-1 stacked honeycomb lattice in $Ba_2Ni(PO_4)_2$, Phys. Rev. B **95**, 024401 (2017).
[5] H. Kawamura, Z2-vortex order of frustrated Heisenberg antiferromagnets in two dimensions, J. Phys.: Conf. Ser. **320**, 012002 (2011).
[6] A. K. Bera, S. M. Yusuf, and S. Banerjee, Short-range magnetic ordering in the geometrically frustrated layered compound $YBaCo_4O_7$ with an extended Kagomé structure, Solid State Sci. **16**, 57 (2013).
[7] S. Fujiki and D. Betts, High Temperature Series Expansion of the Fluctuation of the Vector Chirality for the Spin-1/2 XY Antiferromagnet on the Triangular Lattice, J. Phys. Soc. Jpn. **60**, 435 (1991).
[8] E. A. Zvereva, A.I. Presniakov, H. M. Whangbo, J. H. Koo, V. T. Frantsuzenko, A. O. Savelieva, A. V. Sobolev, B. V. Nalbandyan, S. P. Shih, and J. Chiang, Crucial Role of Site Disorder and Frustration in Unusual Magnetic Properties of Quasi-2D Triangular Lattice Antimonate $Na_4FeSbO_6$, Appl. Magn. Reson. **46**, 1121 (2015).

[9] M. S. Williams, J. P. West, and S. J. Hwu, $KMn_3O_2(Ge_2O_7)$: An S=2 Magnetic Insulator Featuring Pillared Kagome Lattice, Chem. Mater. **26**, 1502 (2014).
[10] R. Moessner and A. Ramirez, Geometrical frustration, Phys. Today **59**, 24 (2006).
[11] D. Schmalfuß, P. Tomczak, J. Schulenburg, and J. Richter, The spin-1/2 Heisenberg antiferromagnet on a 1/7-depleted triangular lattice: Ground-state properties, Phys. Rev. B **65**, 224405 (2002).
[12] Y. Haraguchi, A. Matsuo, K. Kindo, and Z. Hiroi, Frustrated magnetism of the maple-leaf-lattice antiferromagnet $MgMn_3O_7 \cdot 3H_2O$, Phys. Rev. B **98**, 064412 (2018).

[13]	D. J. Farnell, R. Darradi, R. Schmidt, and J. Richter, Spin-half Heisenberg antiferromagnet on two archimedian lattices: From the bounce lattice to the maple-leaf lattice and beyond, Phys. Rev. B **84**, 104406 (2011).
[14]	D. Grohol, K. Matan, J. H. Cho, S. H. Lee, J. W. Lynn, D. G. Nocera, and Y. S. Lee, Spin chirality on a two-dimensional frustrated lattice, Nat. Mater. **4**, 323 (2005).
[15]	J. Schulenburg, J. Richter, and D. Betts, Heisenberg antiferromagnet on a 1/7-depleted triangular lattice, Acta Phys. Pol. A **97**, 971 (2000).
[16]	B. Song, M. Tang, E. Hu, J. O. Borkiewicz, M. K. Wiaderek, Y. Zhang, D. N. Phillip, X. Liu, Z. Shadike and C. Li *et al.*, Understanding the Low-Voltage Hysteresis of Anionic Redox in $Na_2Mn_3O_7$, Chem. Mater. **31**, 3756 (2019).
[17]	K. Sada and P. Barpanda, Layered Sodium Manganese Oxide $Na_2Mn_3O_7$ as an Insertion Host for Aqueous Zinc-ion Batteries, MRS Adv. **4**, 2651 (2019).
[18]	Y. Li, X. Wang, Y. Gao, Q. Zhang, G. Tan, Q. Kong, S. Bak, G. Lu, X. Yang, and L. Gu, Native vacancy enhanced oxygen redox reversibility and structural robustness, Adv. Energy Mater. **9**, 1803087 (2019).

[19]	K. Sada, B. Senthilkumar, and P. Barpanda, Layered $Na_2Mn_3O_7$ as a 3.1 V Insertion Material for Li-Ion Batteries, Appl. Energy Mater. **1**, 6719 (2018).
[20]	Z. Zhang, D. Wu, X. Zhang, X. Zhao, H. Zhang, F. Ding, Z. Xie, and Z. Zhou, First-principles computational studies on layered $Na_2Mn_3O_7$ as a high-rate cathode material for sodium ion batteries, J. Mater. Chem. A **5**, 12752 (2017).
[21]	E. Adamczyk and V. Pralong, $Na_2Mn_3O_7$: A Suitable Electrode Material for Na-Ion Batteries?, Chem. Mater. **29**, 4645 (2017).
[22]	E. Raekelboom, A. Hector, J. Owen, G. Vitins, and M. Weller, Syntheses, structures, and preliminary electrochemistry of the layered lithium and sodium manganese (IV) oxides, $A_2Mn_3O_7$, Chem. Mater. **13**, 4618 (2001).
[23]	F. M. Chang and M. Jansen, Darstellung und Kristallstruktur von $Na_2Mn_3O_7$, Z. Anorg. Allg. Chem. **531**, 177 (1985).
[24]	B. Boisse, S. Nishimura, E. Watanabe, L. Lander, A. Tsuchimoto, J. Kikkawa, E. Kobayashi, D. Asakura, M. Okubo, and A. Yamada, Highly Reversible Oxygen-Redox Chemistry at 4.1 V in $Na_{4/7-x}[\Box_{1/7}Mn_{6/7}]O_2$ ($\Box$: Mn Vacancy), Adv. Energy Mater. **8**, 1800409 (2018).

[25]	C. Venkatesh, B. Bandyopadhyay, A. Midya, K. Mahalingam, V. Ganesan, and P. Mandal, Magnetic properties of the one-dimensional S=3/2 Heisenberg antiferromagnetic spin-chain compound $Na_2Mn_3O_7$, Phys. Rev. B **101**, 184429 (2020).
[26]	J. A. M. Paddison, J. R. Stewart, and A. L. Goodwin, SPINVERT: A program for refinement of paramagnetic diffuse scattering data, J. Phys.: Condens. Matter 25, 454220 (2013).
[27]	B. Saha, A. K. Bera, and S. M. Yusuf, Mechanism of Na-Ion Conduction in the Highly Efficient Layered Battery Material $Na_2Mn_3O_7$, ACS Appl. Energy Mater. **4**, 6040 (2021).
[28]	M. Casas-Cabanas, M. Reynaud, J. Rikarte, P. Horbach, and J. Rodríguez-Carvajal, FAULTS: a program for refinement of structures with extended defects, J. Appl. Crystallogr. **49**, 2259 (2016).
[29]	S. Taniguchi, T. Nishikawa, Y. Yasui, Y. Kobayashi, M. Sato, T. Nishioka, M. Kontani, and K. Sano, Spin Gap Behavior of S= 1/2 Quasi-Two-Dimensional System $Ca_4VO_9$, J. Phys. Soc. Jpn. **64**, 2758 (1995).
[30]	H. Iwase, M. Isobe, Y. Ueda, and H. Yasuoka, Observation of spin gap in $CaV_2O_5$ by NMR, J. Phys. Soc. Jpn. **65**, 2397 (1996).
[31]	K. Ghoshray, B. Pahari, B. Bandyopadhyay, R. Sarkar, and A. Ghoshray, $^{51}$V NMR study of the quasi-one-dimensional alternating chain compound $BaCu_2V_2O_8$, Phys. Rev. B **71**, 214401 (2005).
[32]	B. Pahari, K. Ghoshray, R. Sarkar, B. Bandyopadhyay, and A. Ghoshray, NMR study of $^{51}$V in quasi-one-dimensional integer spin chain compound $SrNi_2V_2O_8$, Phys. Rev. B **73**, 012407 (2006).
[33]	Y. Uchiyama, Y. Sasago, I. Tsukada, K. Uchinokura, A. Zheludev, T. Hayashi, N. Miura, and P. Böni, Spin-vacancy-induced long-range order in a new haldane-gap antiferromagnet, Phys. Rev. Lett. **83**, 632 (1999).
[34]	A. A. Tsirlin, R. Nath, J. Sichelschmidt, Y. Skourski, C. Geibel, and H. Rosner, Frustrated couplings between alternating spin-1 2 chains in $AgVOAsO_4$, Phys. Rev. B **83**, 144412 (2011).

[35] C. Delmas, G. Le Flem, C. Fouassier, and P. Hagenmuller, Etude comparative des proprietes magnetiques des oxydes lamellaires $ACrO_2$ (A= Li, Na, K)—II: Calcul des integrales d'echange, J. Phys. Chem. Solids **39**, 55 (1978).

[36] J. M. Law, H. Benner, and R. K. Kremer, Padé approximations for the magnetic susceptibilities of Heisenberg antiferromagnetic spin chains for various spin values, J. Phys.: Condens. Matter 25, 065601 (2013).

[37] J. Lynn, T. Clinton, W. Li, R. Erwin, J. Liu, K. Vandervoort, and R. Shelton, 2D and 3D magnetic behavior of Er in $ErBa_2Cu_3O_7$, Phys. Rev. Lett. **63**, 2606 (1989).

[38] H. Zhang, J. Lynn, and D. Morris, Coupled-bilayer two-dimensional magnetic order of the Dy ions in $Dy_2Ba_4Cu_7O_{15}$, Phys. Rev. B **45**, 10022 (1992).

[39] S. M. Yusuf, A. K. Bera, N. S. Kini, I. Mirebeau, and S. Petit, Two-and three-dimensional magnetic correlations in the spin-1/2 square-lattice system $Zn_2VO(PO_4)_2$, Phys. Rev. B **82**, 094412 (2010).

[40] S. M. Yusuf, J. M. De Teresa, P. A. Algarabel, M. D. Mukadam, I. Mirebeau, J. M. Mignot, C. Marquina, and M. R. Ibarra, Two- and three-dimensional magnetic ordering in the bilayer manganite $Ca_{2.5}Sr_{0.5}GaMn_2O_8$, Phys. Rev. B **74**, 184409 (2006).

[41] A. K. Bera, S. M. Yusuf, and I. Mirebeau, Effect of electron doping on the magnetic correlations in the bilayered brownmillerite compound Ca2.5−xLaxSr0.5GaMn2O8: A neutron diffraction study, J. Phys.: Condens. Matter 23, 426005 (2011).

[42] A. K. Bera, S. M. Yusuf, A. Kumar, and C. Ritter, Zigzag antiferromagnetic ground state with anisotropic correlation lengths in the quasi-two-dimensional honeycomb lattice compound $Na_2Co_2TeO_6$, Phys. Rev. B **95**, 094424 (2017).

[43] A. K. Bera, S. M. Yusuf, L. Keller, F. Yokaichiya, and J. R. Stewart, Magnetism of two-dimensional honeycomb layered $Na_2Ni_2TeO_6$ driven by intermediate Na-layer crystal structure, Phys. Rev. B **105**, 014410 (2022).

[44] A. K. Bera, S. M. Yusuf, A. Kumar, M. Majumder, K. Ghoshray, and L. Keller, Long-range and short-range magnetic correlations, and microscopic origin of net magnetization in the spin-1 trimer chain compound $CaNi_3P_4O_{14}$, Phys. Rev. B **93**, 184409 (2016).

[45] J. A. Paddison and A. L. Goodwin, Empirical magnetic structure solution of frustrated spin systems, Phys. Rev. Lett. **108**, 017204 (2012).

[46] G. J. Nilsen, C. M. Thompson, G. Ehlers, C. A. Marjerrison, and J. E. Greedan, Diffuse magnetic neutron scattering in the highly frustrated double perovskite $Ba_2YRuO_6$, Phys. Rev. B **91**, 054415 (2015).

[47] J. A. Paddison, S. Agrestini, M. R. Lees, C. L. Fleck, P. P. Deen, A. L. Goodwin, J. R. Stewart, and O. A. Petrenko, Spin correlations in $Ca_3Co_2O_6$: polarized-neutron diffraction and Monte Carlo study, Phys. Rev. B **90**, 014411 (2014).

[48] S. Siebeneichler, K. V. Dorn, A. Ovchinnikov, W. Papawassiliou, I. da Silva, V. Smetana, A. J. Pell, and A.-V. Mudring, Frustration and 120° Magnetic Ordering in the Layered Triangular Antiferromagnets $AFe(PO_3F)_2$ (A= K,$(NH_4)_2$Cl, $NH_4$, Rb, and Cs), Chem. Mater. **34**, 7982 (2022).

[49] H. Serrano-González, S. Bramwell, K. Harris, B. Kariuki, L. Nixon, I. Parkin, and C. Ritter, Magnetic structures of the triangular lattice magnets $AFe(SO_4)_2$ (A= K, Rb, Cs), J. Appl. Phys. **83**, 6314 (1998).

[50] T. Inami, Neutron powder diffraction experiments on the layered triangular-lattice antiferromagnets $RbFe(MoO_4)_2$ and $CsFe(SO_4)_2$, J. Solid State Chem. **180**, 2075 (2007).

[51] A. B. Harris, C. Kallin, and A. J. Berlinsky, Possible Néel orderings of the Kagomé antiferromagnet, Phys. Rev. B **45**, 2899 (1992).

[52] T. Inami, M. Nishiyama, S. Maegawa, and Y. Oka, Magnetic structure of the kagomé lattice antiferromagnet potassium jarosite $KFe_3(OH)_6(SO_4)_2$, Phys. Rev. B **61**, 12181 (2000).

[53] D. Kozlenko, A. F. Kusmartseva, E. Lukin, D. Keen, W. Marshall, M. De Vries, and K. V. Kamenev, From quantum disorder to magnetic order in an S=1/2 kagome lattice: A structural and magnetic study of herbertsmithite at high pressure, Phys. Rev. Lett. **108**, 187207 (2012).

[54] W.-m. Zhang, W. Saslow, and M. Gabay, Row generalization of the fully frustrated triangular XY model, Phys. Rev. B **44**, 5129 (1991).

[55] A. I. Smirnov, L. E. Svistov, L. A. Prozorova, A. Zheludev, M.D. Lumsden, E. Ressouche, O. A. Petrenko, K. Nishikawa, S. Kimura, and M. Hagiwara, Chiral and collinear ordering in a distorted triangular antiferromagnet, Phys. Rev. Lett. **102**, 037202 (2009).

[56] I. Tanaka and H. Tsunetsugu, Nematicity Liquid in a Trimerized-Kagome Antiferromagnet, J. Phys. Soc. Jpn. **90**, 063707 (2021).

[57] Y. Iqbal, D. Poilblanc, R. Thomale, and F. Becca, Persistence of the gapless spin liquid in the breathing kagome Heisenberg antiferromagnet, Phys. Rev. B **97**, 115127 (2018).

[58] F. Wang, A. Vishwanath, and Y. B. Kim, Quantum and classical spins on the spatially distorted kagomé lattice: Applications to volborthite $CuV_2O_7(OH)_2 \cdot 2H_2O$, Phys. Rev. B **76**, 094421 (2007).

[59] W. Li, S.-S. Gong, Y. Zhao, S.-J. Ran, S. Gao, and G. Su, Phase transitions and thermodynamics of the two-dimensional Ising model on a distorted kagome lattice, Phys. Rev. B **82**, 134434 (2010).

[60] M. Hering, F. Ferrari, A. Razpopov, I. I. Mazin, R. Valentí, H. O. Jeschke, and J. Reuther, Phase diagram of a distorted kagome antiferromagnet and application to Y-kapellasite, npj Comput. Mater. **8**, 1 (2022).

[61] H. Lu and H. Kageyama, $PbFePO_4F_2$ with a 1/6th bond depleted triangular lattice, Dalton Trans. **47**, 15303 (2018).

**Figures**

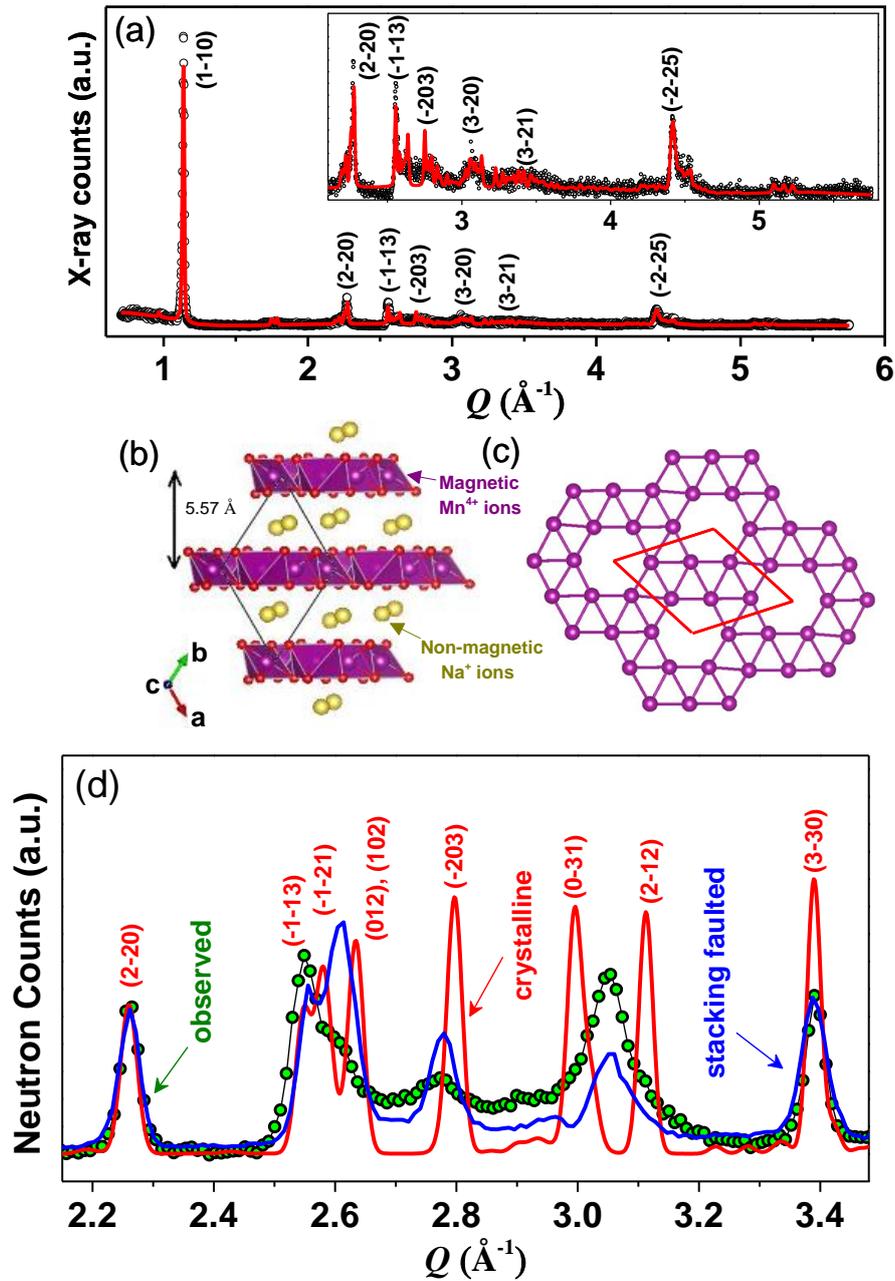

**Fig. 1:** (a) The powder x-ray diffraction pattern at room temperature. The black scattered points are observed experimental data and the continuous red line is the profile fitted curve (sample annealed for 4 hours). The inset shows the magnified view of the broad diffuse Bragg peaks. (b) The schematic layered crystal structure of $Na_2Mn_3O_7$. The red solid line represents the unit cell dimension. (c) The lattice geometry of the $Mn^{4+}$ ions ($S=3/2$) within the 2D magnetic planes. (d) The powder neutron diffraction pattern (symbols) at room temperature for $Na_2Mn_3O_7$ (sample annealed for 4 hours). The red and blue curves are the calculated diffraction patterns by considering the instrument $Q$-resolution parameters for purely crystalline (ICSD_collcode5665) and stacking faulted structure, respectively.

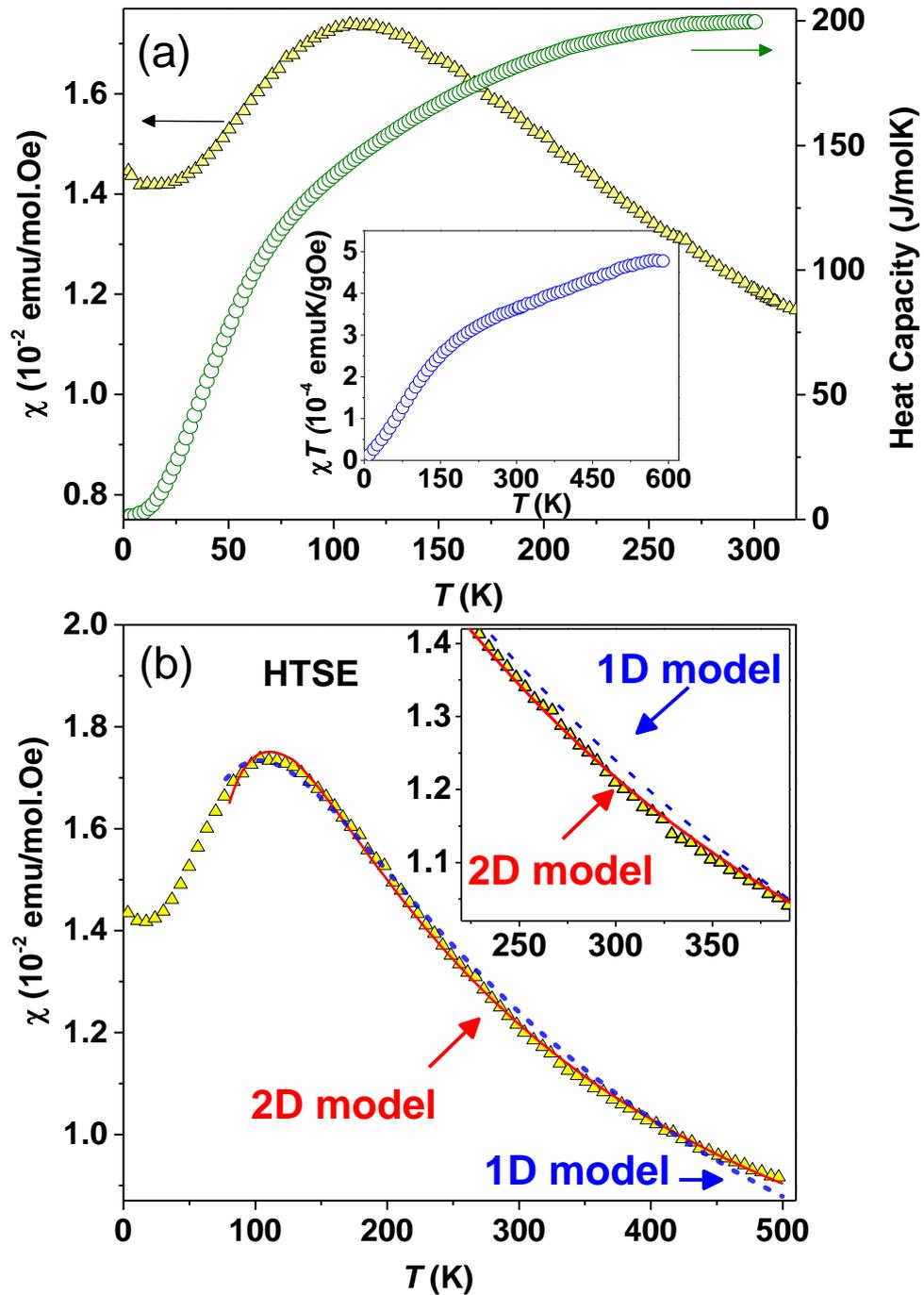

**Fig. 2:** (a) Temperature dependent dc-magnetic susceptibility ($\chi=M/H$), measured under $H$=10 kOe, and zero-field heat capacity curves of Na$_2$Mn$_3$O$_7$. Inset shows the $\chi T$ vs $T$ plot revealing the non-paramagnetic type behaviour. (b) The $\chi(T)$ curve (measured under 10 kOe) over 2-500 K. The red solid and blue dotted lines are the fitted curves for 2D and 1D models, respectively. The inset shows a magnified view of $\chi(T)$ curve over the selected temperature range, revealing a better fitting for the 2D model.

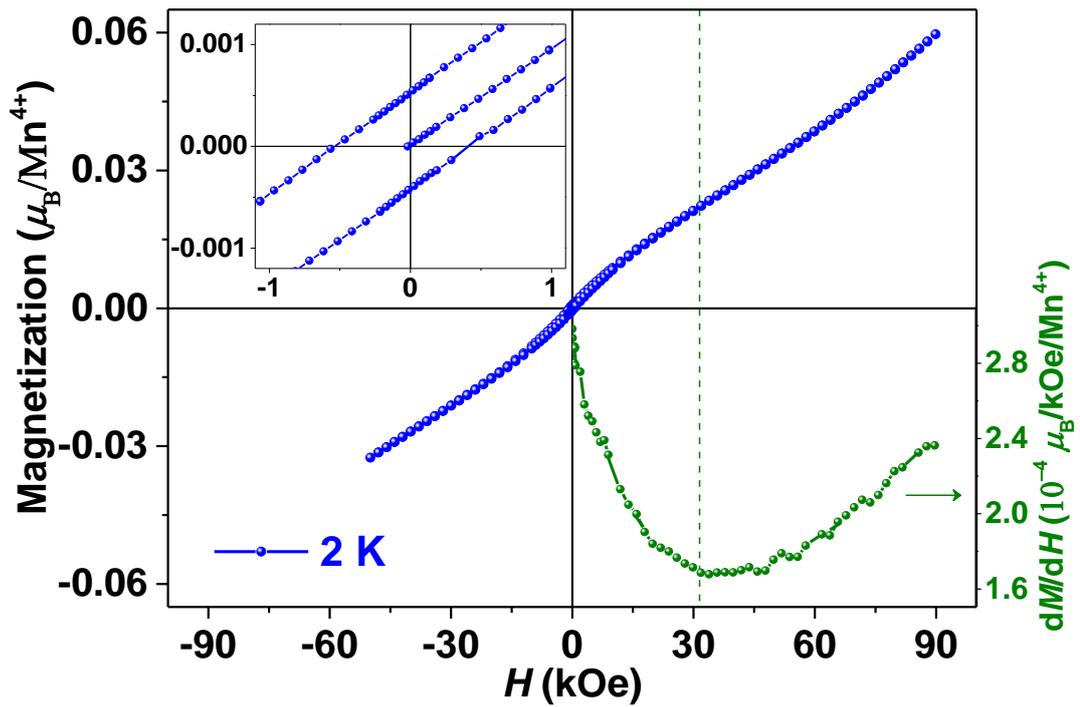

**Fig. 3:** Isothermal magnetization curves, measured at 2 K, for $Na_2Mn_3O_7$. The top-left inset shows an enlarged view of isothermal magnetization over low magnetic field region. The bottom-right inset shows the derivative curve ($dM/dH$ vs $H$) revealing a field induced transition at ~ 30 kOe.

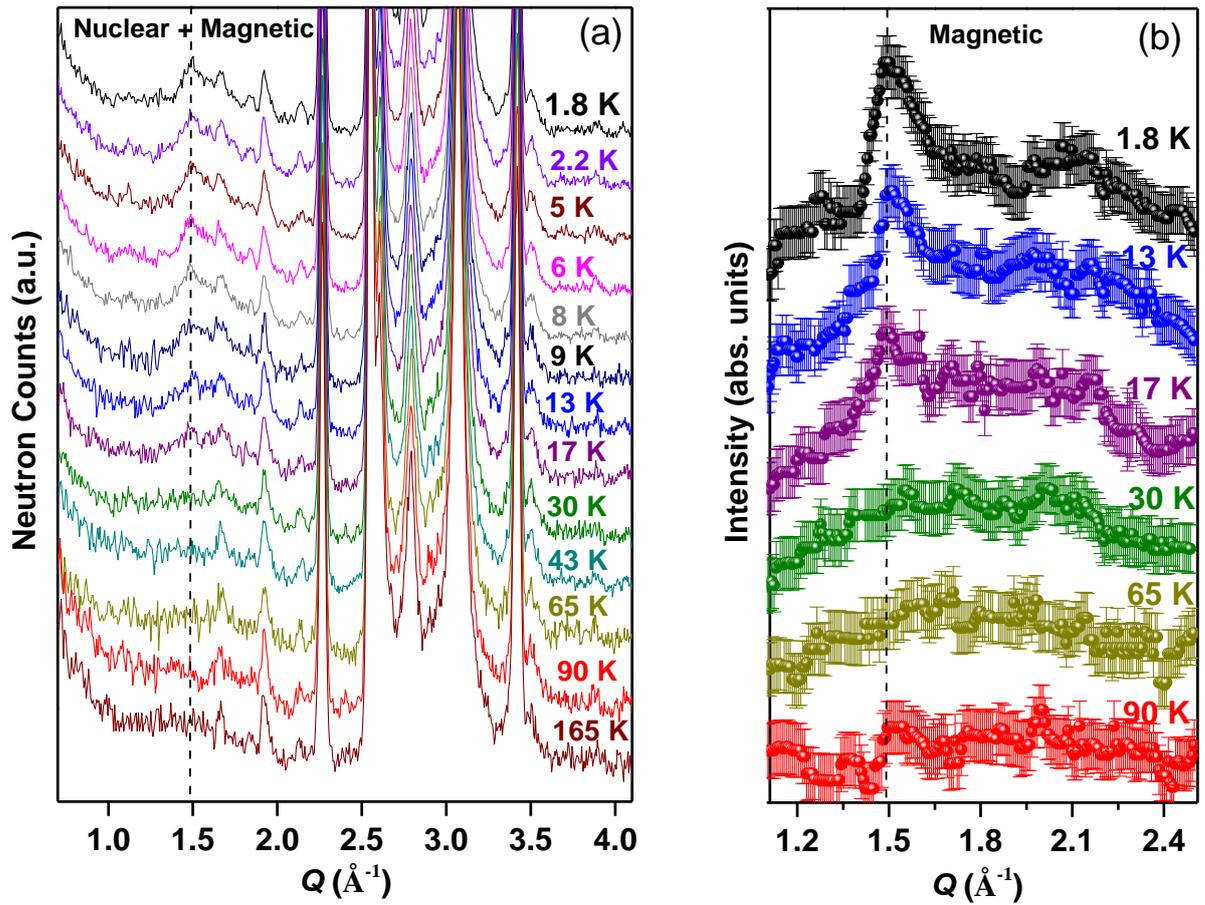

**Fig. 4:** (a) As measured neutron diffraction patterns over 1.8-165 K. (b) The pure magnetic diffraction patterns after subtraction of paramagnetic background (measured at 165 K). Data were recorded using the E6 diffractometer.

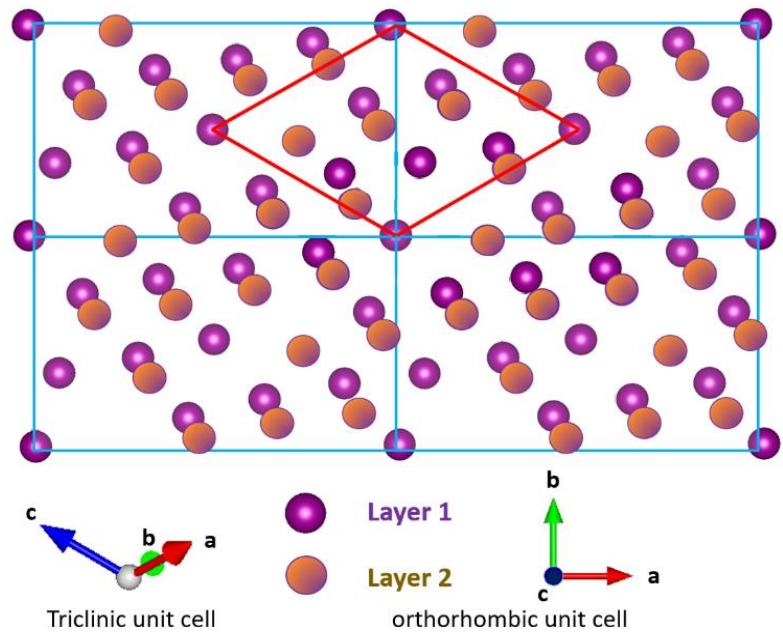

**Fig. 5:** The local crystal structure of $Na_2Mn_3O_7$ showing the magnetic $Mn^{4+}$ ions alone. Only successive Manganese-layers are shown for clarity. The triclinic unit cell is shown by red line. The transformed orthorhombic unit cell used for *RMC* analysis is shown by blue lines.

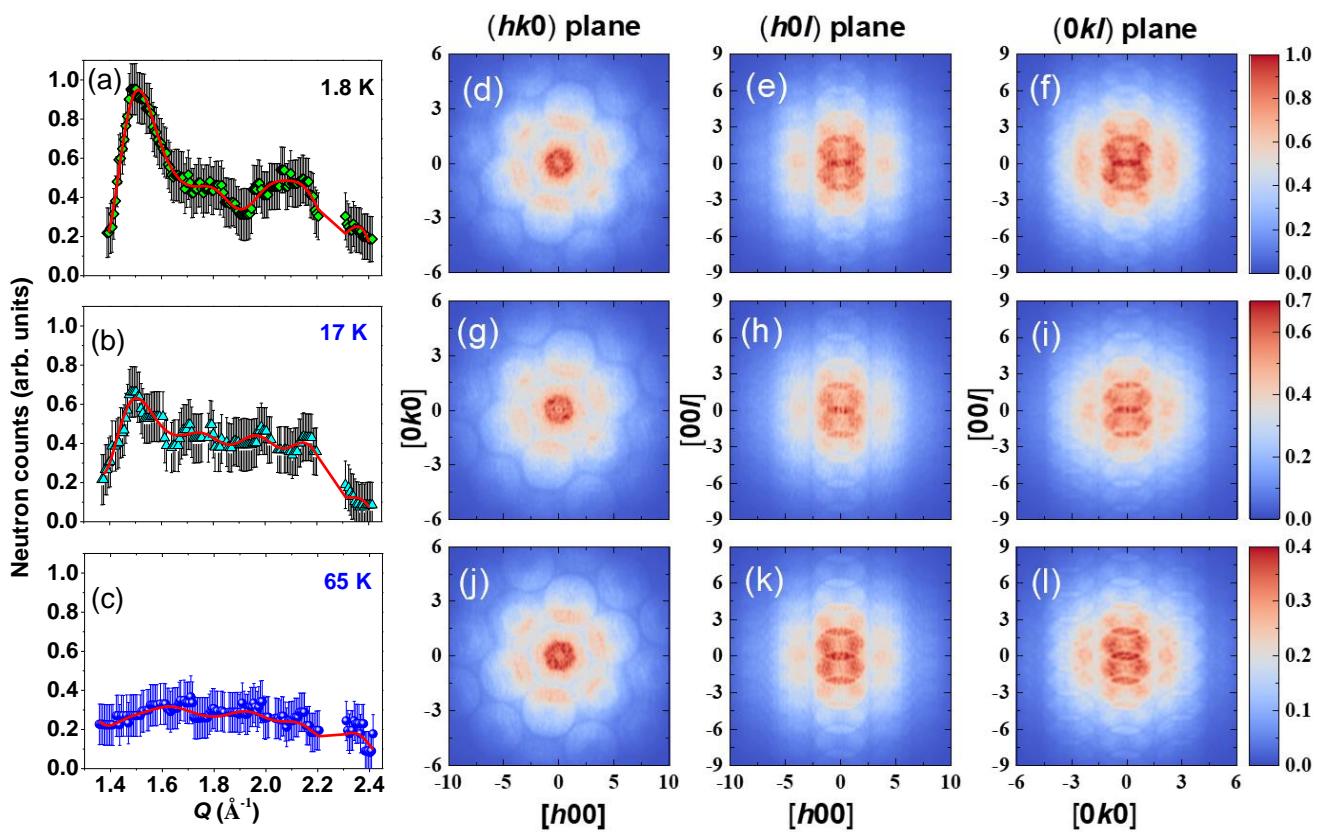

**Fig. 6:** (a-c) The experimental powder magnetic diffuse scattering patterns at lower temperatures 1.8, 17 and 65 K after subtracting the paramagnetic background (pattern measured at 165 K). The solid red lines are the calculated scattering patterns by the RMC method. (d-l) The reconstructed diffuse magnetic scattering patterns in selected scattering planes (*hk0*), (*h0l*) and (*0kl*), respectively, corresponding to (a), (b) and (c). The intensity variation is shown by the colour scheme (blue to red for minimum to maximum intensity).

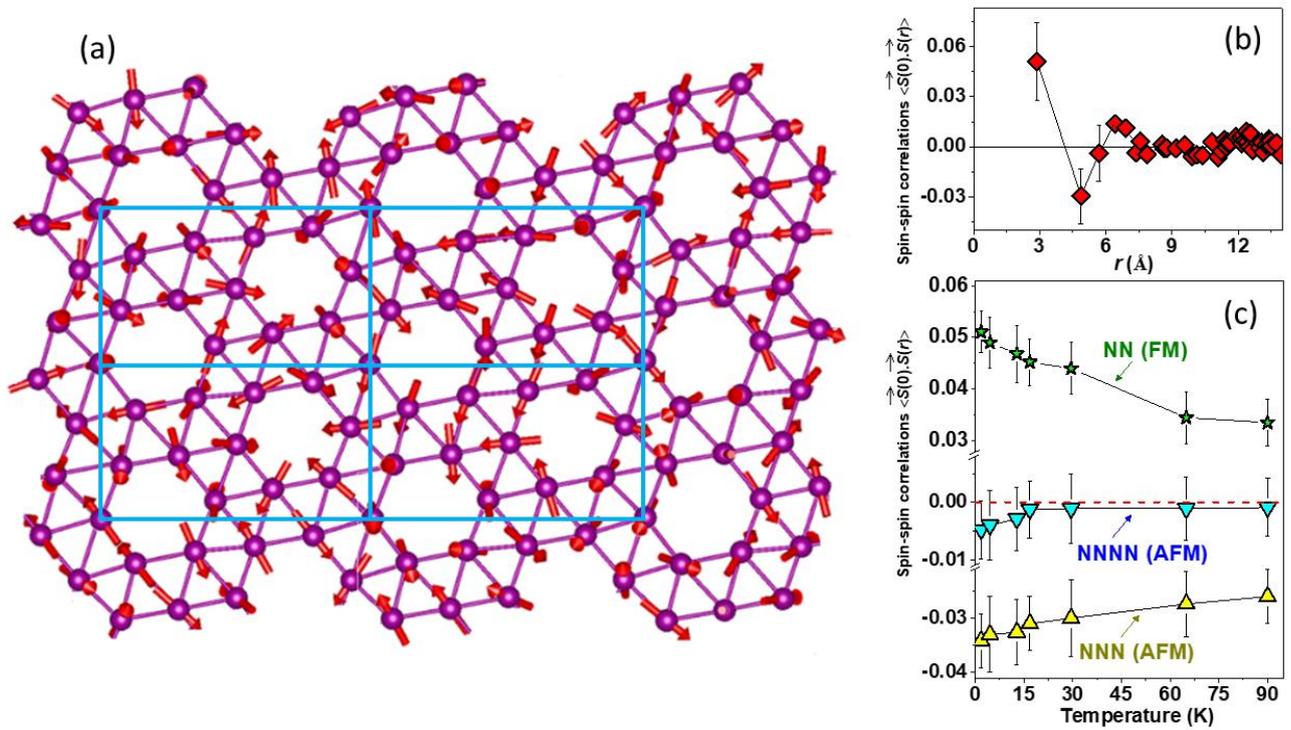

**Fig. 7:** (a) Part of a spin-configuration solution obtained from a RMC fitting of the powder magnetic diffuse scattering pattern at 1.8 K. (b) The real space spin-pair correlations in $Na_2Mn_3O_7$ at 1.8 K (averaged over spin-configuration solutions of 15 independent RMC refinements). (c) Temperature dependent in-plane (within MLL) NN, NNN, and NNNN spin-spin correlations.

Page **22** of 26

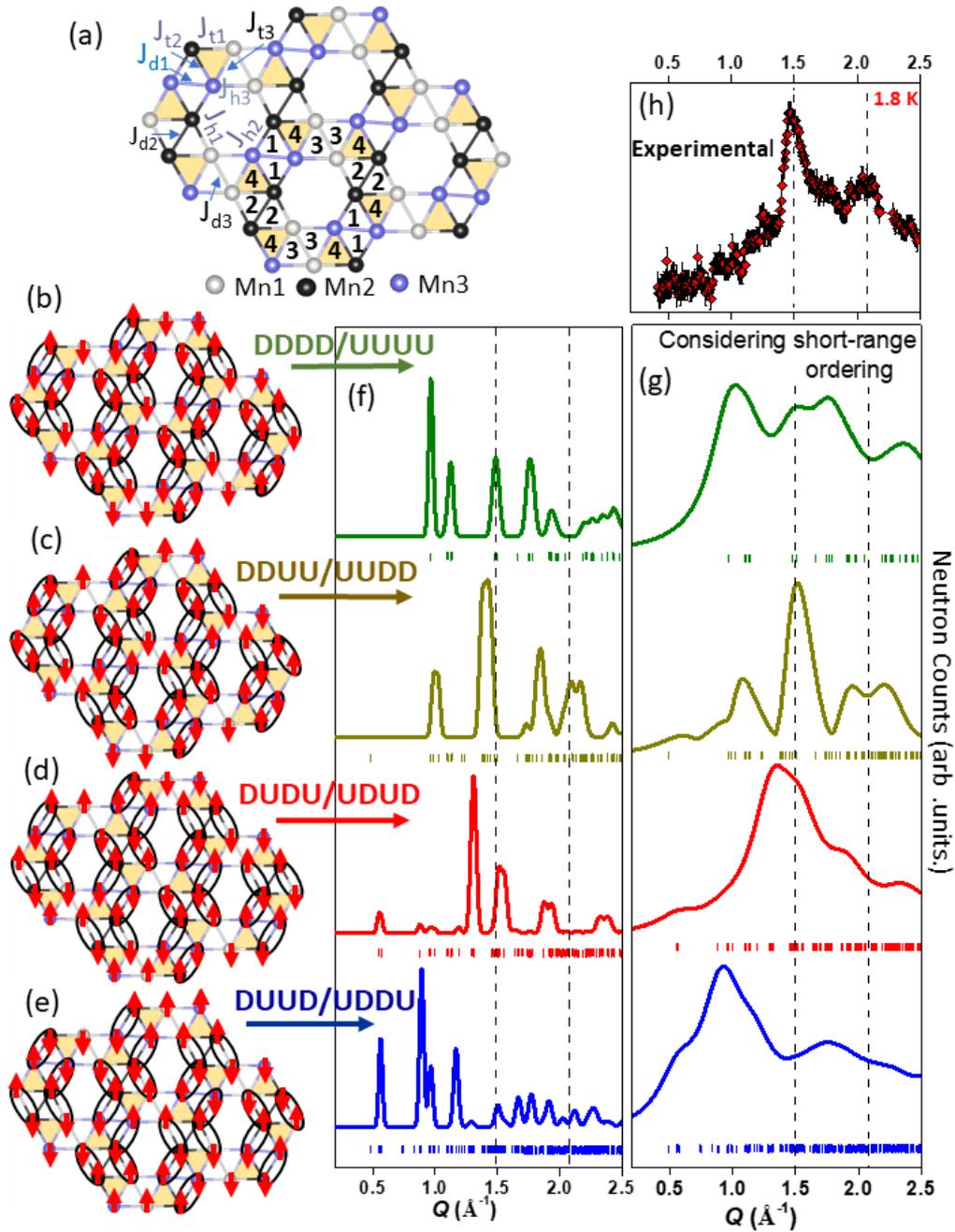

**Fig. 8:** (a) The MLL lattice of $Na_2Mn_3O_7$ with nine non-equivalent exchange interactions. The non-equivalent exchange interactions lead to four different types of triangles (labelled as 1-4) within the hexagonal unit of the MLL. Possible ground state magnetic structures (b) DDDD/UUUU, (c) DDUU/UUDD, (d) DUDU/UDUD, and (e) DUUD/UDDU for the distorted MLL in $Na_2Mn_3O_7$. The bonds marked by the solid black ovals represent the AFM exchange interactions. The arrows represent the spins associated with each magnetic atom. (f) The simulated magnetic neutron diffraction patterns for the respective magnetic ground state structures. (g) The simulated magnetic neutron diffraction patterns considering short-range magnetic correlations. (h) The experimentally measured magnetic diffuse scattering patterns at 1.8 K.

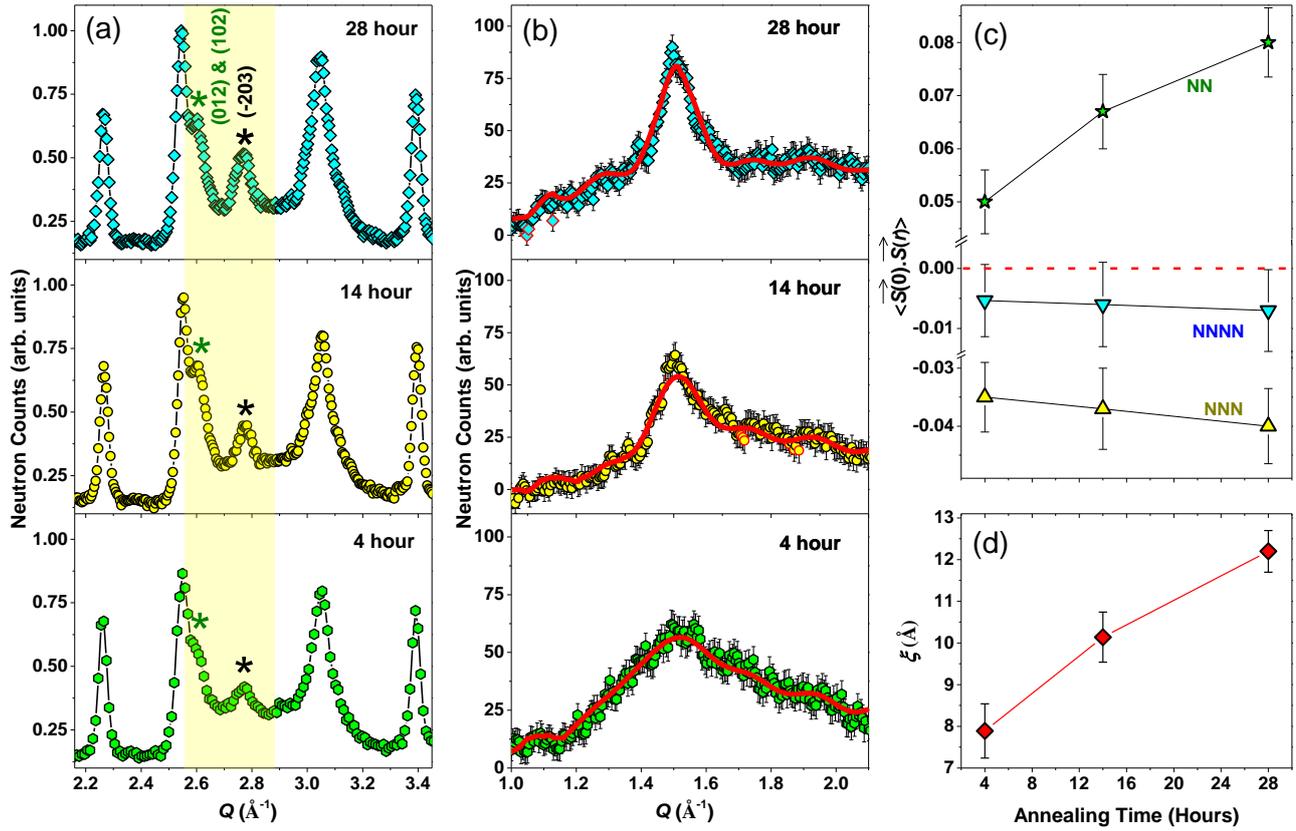

**Fig. 9:** (a) Experimental neutron diffraction patterns, measured at 300 K using the PD-I, of $Na_2Mn_3O_7$ prepared with different annealing times of 4, 14, and 28 hours (bottom to top panel), respectively. The diffraction patterns are normalized with respect to the Bragg peak (situated over $Q \sim 2.25$ Å$^{-1}$) which is unaffected by stacking faults. Substantial changes in the selective Bragg peaks are marked with asterisks indicating the increase in crystallinity (reduction of stacking fault) with annealing time. (b) The pure magnetic neutron diffraction patterns at 5 K measured using the PD-I (after subtraction of the paramagnetic background measured at 100 K) for $Na_2Mn_3O_7$ samples synthesized with different annealing times of 4, 14 and 28 hours. The solid red lines are the calculated patterns by the RMC method. The variation of (c) real space spin-spin correlation function [$\langle \vec{S}(0)\cdot\vec{S}(r)\rangle$], and (d) spin-spin correlation length ($\xi$) with the annealing time.

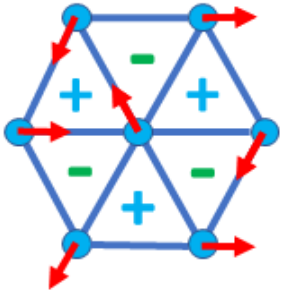
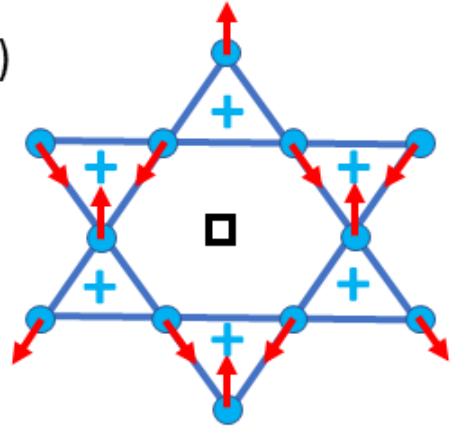
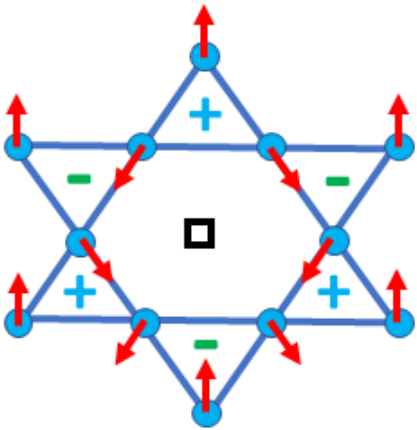
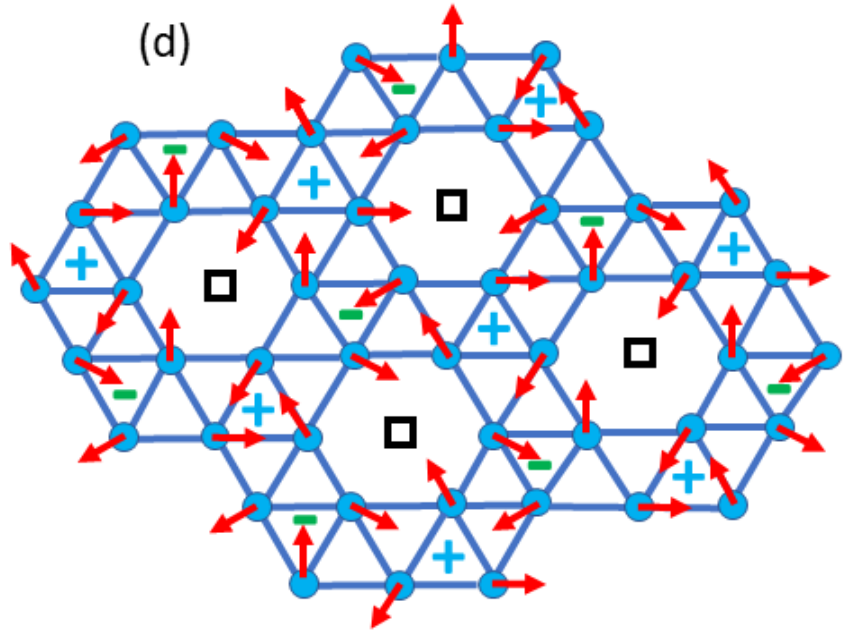

**Fig. 10:** Geometrically frustrated 2D magnetic lattices (a) triangular, (b-c) kagome, and (d) maple leaf lattices along with their chiral spin structures with NN exchange interactions. The square symbols show the missing lattice points for the kagome and maple leaf lattices.

Table I: Bond length and bond angles of $Na_2Mn_3O_7$ (considering the standard structure ICSD database code ICSD5665).

| Interaction | Magnetic ions | Mn-Mn Direct Distance (Å) | Pathways | Bond Lengths | Bond Angles (deg) |
|---|---|---|---|---|---|
| $J_{d1}$ | Mn3-Mn3 | 2.923(7) | Mn3-O5-Mn3 | Mn3-O5=1.938(3)<br>O5- Mn3=1.949(2) | 97.5(10) |
| $J_{d2}$ | Mn2-Mn2 | 2.974(3) | Mn2-O6-Mn2 | Mn2-O6=1.954(9)<br>O6-Mn2=1.949(2) | 99.3(6) |
| $J_{d3}$ | Mn1-Mn1 | 2.968(6) | Mn1-O3-Mn1 | Mn1-O3=1.957(14)<br>O3- Mn1=2.039(11) | 95.9(6) |
| $J_{t1}$ | Mn2-Mn1 | 2.876(5) | Mn2-O6-Mn1<br>Mn2-O2-Mn1 | Mn2-O6=1.954(3)<br>O6- Mn1=1.905(13)<br>Mn2-O2=1.924(14)<br>O2- Mn1=1.953(11) | 96.3(5)<br>95.8(6) |
| $J_{t2}$ | Mn3-Mn2 | 2.930(5) | Mn3-O2-Mn2<br>Mn3-O5-Mn2 | Mn2-O2=2.011(11)<br>O2- Mn3=1.924(14)<br>Mn2-O5=1.942(11)<br>O5- Mn3=1.949(14) | 96.2<br>97.7(6) |
| $J_{t3}$ | Mn3-Mn1 | 2.939(3) | Mn3-O2-Mn1<br>Mn3-O3-Mn1 | Mn3-O2=2.011(11)<br>O2- Mn1=1.953(11)<br>Mn3-O3=1.919(11)<br>O3- Mn1=2.039(11) | 95.7(5)<br>95.9(6) |
| $J_{h1}$ | Mn1-Mn1 | 2.968(6) | Mn1-O3-Mn1 | Mn1-O3=1.957(14)<br>O3- Mn1=2.039(11) | 95.9(6) |
| $J_{h1}$ | Mn2-Mn1 | 2.783(5) | Mn2-O4-Mn1<br>Mn2-O6-Mn1 | Mn2-O4=1.897(13)<br>O4- Mn1=1.842(11)<br>Mn2-O6=1.949((10)<br>O6- Mn1=1.905(13) | 96.2(6) |
| $J_{h2}$ | Mn3-Mn2 | 2.726(3) | Mn3-O5-Mn2<br>Mn3-O7-Mn2 | Mn3-O5=1.938(13)<br>O5- Mn2=1.942(11)<br>Mn3-O7=1.814(9)<br>O7- Mn2=1.836(10) | 89.2(6)<br>96.6(6) |
| $J_{h3}$ | Mn3-Mn1 | 2.865(5) | Mn3-O1-M1<br>Mn3-O3-M1 | Mn3-O1=1.798(13)<br>O1- Mn1=1.896(12)<br>Mn3-O3=1.919(11)<br>O3- Mn1=1.957(14) | 101.7(6)<br>95.4(5) |